\documentclass{article}

\usepackage[a4paper,top=2cm,bottom=2cm,left=2cm,right=2cm]{geometry}
\usepackage{enumitem}
\usepackage{csquotes}

\usepackage[
  backend=biber,
  style=apa,
  sorting=nyt
]{biblatex}

\usepackage{longtable}
\usepackage{booktabs}
\usepackage{graphicx}
\usepackage{rotating}
\usepackage{array}
\usepackage{float}
\usepackage{placeins}
\usepackage{xurl}
\usepackage[hidelinks]{hyperref}
\usepackage{tabularx}
\usepackage{authblk}
\usepackage{orcidlink}
\title{Detecting Hallucinated and Suspicious Citations: What Current Tools Can and Cannot Do}

\author[1]{Fidan Badalova\,\orcidlink{0009-0005-3525-0717}}
\author[1]{Philipp Mayr\,\orcidlink{0000-0002-6656-1658}}

\affil[1]{GESIS -- Leibniz Institute for the Social Sciences, Cologne, Germany}

\affil[*]{Corresponding author: Fidan Badalova, 
\href{mailto:fidan.badalova@gesis.org}{fidan.badalova@gesis.org}}
\date{}
\begin{document}

\maketitle

\begin{abstract}

Large language models are increasingly used in academic writing, including for reference generation, raising concerns about hallucinated and unreliable citations. Recent research suggests that this problem is already widespread and is becoming increasingly prevalent in the published literature and at scientific conferences. In this position paper, we review recent studies on hallucinated references and evaluate several currently available tools for detecting problematic references using documents containing hallucinated citations. The tools assessed include CheckIfExist, HalluCiteChecker, Hallucinator, Hallucinated Reference Finder (HalRef), and RefChecker. While these systems can provide useful early warnings in many cases, their performance is limited by reference extraction errors, incomplete metadata, limited database coverage, and inconsistent verification results. We argue that hallucinated and suspicious references have become a real and growing problem for scientific communication, and that more transparent and multi-source detection systems are still needed.

\end{abstract}
\noindent\textbf{Keywords:} hallucinated citations; reference verification; scholarly communication; citation integrity

\section{Introduction}
In recent years, particularly following the development and widespread adoption of large language models (LLMs), these models have been increasingly used in academic writing, especially for literature searches and reference generation. These generated references may be hallucinated, fabricated, or bibliographically problematic, posing an increasingly serious risk to scholarly communication \parencite{farquhar2024semanticentropy, gibney2025aicitations,  jain2025citation, lee2026reflens}. This problem is not limited to individual technical errors; it also encompasses a broader range of issues, including the presentation of non-existent sources as genuine references, the use of incorrect DOIs, inconsistencies in author names, titles, and publication venue information, and the inclusion of references in scholarly texts that appear convincing but cannot actually be verified \parencite{chen2026fake,  emsley2023fabrications,spennemann2026unique}. 

The importance of this issue is already evident in both the scientific publications and academic conferences. For example, a recent \textit{Nature} analysis reported that tens of thousands of publications from 2025 may contain invalid references generated by artificial intelligence \parencite{naddaf2026hallucinated}. Additional evidence supports this concern. A large-scale audit published in  \textit{The Lancet} reported a substantial increase in fabricated citations in the biomedical literature \parencite{topaz2026fabricated}, while the large-scale study by Zhao and colleagues demonstrated that non-existent references occur systematically across different publication corpora \parencite{zhao2026llmhallucinations}. 
In addition, an earlier study by Walters found that references generated by ChatGPT may be entirely fabricated or contain serious bibliographic errors \parencite{walters2023fabrication}. The detection of such cases in conference settings further demonstrates that this problem is not merely theoretical, but also practical and immediate. For example, in recent months, concrete instances of hallucinated references have been publicly documented in both ICLR 2026 submissions and accepted NeurIPS papers \parencite{gptzero2025iclr, gptzero2026neurips}. ACL 2026 also reported that final camera-ready checks identified citations to non-existent literature in more than 100 accepted papers \parencite{acl2026statement}. Similarly, other studies indicate that this problem is not limited to a single platform or discipline but recurs across diverse scientific contexts and usage scenarios, and is likely to persist \parencite{erdem2025financial, sakai2026hallucitation,watson2024hallucinated, wu2025sourcecheckup}.

From this perspective, the issue is not merely that AI tools sometimes generate incorrect references; the more serious concern is that such references may appear credible at first glance, escape detection by reviewers and editors, and ultimately undermine the reliability of the scientific literature \parencite{topaz2026fabricated, zhao2026llmhallucinations}. This paper discusses both how this problem has been documented in recent studies and the extent to which existing detection tools can identify such cases.

The main position of this article is that hallucinated and suspicious references have become a real and growing threat to scholarly communication. 
Although several open research tools for detecting hallucinated citations are now available, and platforms such as arXiv have started to respond more rigorously to this issue \parencite{chawla2026arxivban}, current mitigation measures remain insufficiently reliable. Recent studies have shown the scale of this problem, but existing verification systems still depend heavily on the quality of reference extraction, database coverage, metadata completeness, and the matching algorithms. Consequently, while these tools are valuable for preliminary screening, their results should not yet be regarded as definitive. We therefore argue for the development of reference-checking systems that are more transparent, reproducible, and capable of multi-source verification.

\section{Evidence that Hallucinated References are a real problem}

The position in this article is supported by evidence from recent studies, comparative evaluations of existing tools, and practical experiments. Recent research demonstrates that hallucinated and suspicious references are a growing challenge for scholarly communication. Comparisons of existing detection tools show that these systems identify problematic references using different approaches and with varying levels of effectiveness. Practical evaluations further indicate that, although these tools are useful for preliminary screening, their results are not always reliable because of errors in citation extraction, metadata inconsistencies, and limitations in database coverage. These findings highlight the need for more robust, transparent, and reliable systems for detecting hallucinated references.

\subsection{Evidence from recent studies}
In Table 1, we compare and describe two recent large-scale studies on fabricated and hallucinated references that have attracted lots of attention this year. %These studies were selected because they examine the issue using large-scale corpora and provide important empirical evidence on the prevalence of problematic references in the scientific literature.
%\begin{itemize}
   % \item The Lancet study of fabricated citations in published biomedical literature and estimates how frequently they occur \parencite{topaz2026fabricated}.
    %\item The multi-corpus study by Zhao et al. on LLM-related hallucinated citations across preprint and publication repositories \parencite{zhao2026llmhallucinations}.
%\end{itemize}

\small 
\begin{longtable}{|p{3cm}|p{5.2cm}|p{5.2cm}|} \caption{Comparison of two studies on fabricated and hallucinated references} \label{tab:comparison_hallucinated_references} \\ 
\hline 
\textbf{Aspect} & \textbf{Study 1: The Lancet \parencite{topaz2026fabricated}} & \textbf{Study 2: The multi-corpus study \parencite{zhao2026llmhallucinations}} \\ \hline \endfirsthead \hline \textbf{Aspect} & \textbf{Study 1: The Lancet } & \textbf{Study 2: Multi-corpus LLM hallucinated citation study} \\ 
\hline 
\endhead 
\hline 
\endfoot 
\hline 
\endlastfoot

Scale &
Approximately 2.5 million papers and 125.6 million references. Approximately 97.1 million references with a DOI or PubMed ID were verified. &
Approximately 2.5 million papers and 111 million references across four corpora. \\
\hline
Data sources &
PubMed Central Open Access articles. &
arXiv, bioRxiv, SSRN, and PubMed Central. \\
\hline
Disciplinary coverage &
Biomedical and medical literature. &
Computer science, physics, mathematics, biology, social sciences, law, humanities, and biomedical literature. \\
\hline
Time period &
January 2023 to February 2026. &
Mainly a comparison between the pre-LLM period and the post-LLM period, especially 2020--2025. \\
\hline

Definition of problematic reference &
A reference was considered suspicious or fabricated if the DOI/PMID metadata did not match the cited bibliographic information and the cited title could not be found in major scholarly databases. &
A reference was considered potentially hallucinated if its title could not be matched in major scholarly databases and if the unmatched-reference rate increased beyond the pre-LLM baseline. \\
\hline
Detection approach &
Identifier-based verification: DOI and PubMed ID resolution, metadata comparison, title verification, filtering of false positives, and manual validation. &
Title-based matching: extraction of reference titles, database matching, comparison with the pre-LLM baseline, and estimation of excess unmatched references after 2023. \\
\hline
Databases used for verification &
PubMed, Crossref, OpenAlex, and Google Scholar. &
Semantic Scholar, OpenAlex, and Google Scholar. \\
\hline
Main methods &
XML reference extraction, DOI/PMID resolution, metadata comparison, text similarity, automated filtering, LLM-assisted screening, and manual validation. &
Reference extraction, title parsing, scholarly database matching, unmatched-reference rate calculation, and pre/post-LLM baseline comparison. \\

\end{longtable}

\normalsize

Both studies suggest that problematic citations are already widespread. The Lancet study \parencite{topaz2026fabricated} found 4,046 fabricated citations in 2,810 biomedical papers and showed that their prevalence increased from 2023 to 2025/2026. 
In 2023, a fabricated reference was identified in approximately one in 2,828 papers; in 2025, this increased to one in 458 papers; and in early 2026, to one in 277 papers. The Zhao et al. study \parencite{zhao2026llmhallucinations} likewise found that hallucinated citations increased particularly after mid-2024 and estimated that around 146,932 such citations may have appeared across the four corpora in 2025. The estimated rates for 2025 were 0.39\% for arXiv, 0.21\% for bioRxiv, 1.91\% for SSRN, and 0.27\% for PubMed Central.

The Lancet study does not prove that every fabricated citation was directly created by an LLM, and it suggests that possible causes include LLM hallucinations, paper mill activity, or intentional manipulation. The Zhao et al. study also interprets the post-2023 increase as a signal consistent with LLM use, but it does not conclude causality at the level of individual citations. At the same time, both studies show that problematic citations can pass editorial or moderation processes. However, limitations such as database coverage, the possibility of false positives, and the existence of real but unindexed citations must also be taken into account.

\subsection{Evidence from tool comparison}

%To better understand the available approaches to detecting hallucinated citations, we contrast five recent tools in Table 2: \textit\textit{CheckIfExist} \parencite{abbonato2026checkifexist}, \textit{HalluCiteChecker} \parencite{sakai2026hallucitechecker}, \textit{Hallucinator} \parencite{hallucinator2026}, \textit{Hallucinated Reference Finder (HalRef)} \parencite{jurgens2026halref}, and  \textit{RefChecker} \parencite{russinovich2026refchecker, russinovich2026phantom}. The five tools compared in this study address problematic references using different approaches, ranging from simple existence checking to PDF-based reference extraction, multi-database matching, and structured reference verification. Their main technical characteristics are summarized in Table 2.

Table 2 compares five recent hallucinated reference detection tools \textit{CheckIfExist} \parencite{abbonato2026checkifexist}, \textit{HalluCiteChecker} \parencite{sakai2026hallucitechecker}, \textit{Hallucinator} \parencite{hallucinator2026}, \textit{Hallucinated Reference Finder (HalRef)} \parencite{jurgens2026halref}, and \textit{RefChecker} \parencite{russinovich2026refchecker, russinovich2026phantom} which range from basic existence checking to multi-database and structured reference verification. 

\begin{table}[H]
\centering
\caption{Comparison of hallucinated citation detection tools}
\label{tab:tool_comparison}
\renewcommand{\arraystretch}{1.2}
\setlength{\tabcolsep}{4pt}
\resizebox{\textwidth}{!}{%
\begin{tabular}{|p{2.7cm}|p{3.8cm}|p{3.8cm}|p{3.8cm}|p{3.8cm}|p{3.8cm}|}
\hline
\textbf{Feature} & \textbf{CheckIfExist} & \textbf{HalluCiteChecker} & \textbf{Hallucinator} & \textbf{HalRef} & \textbf{RefChecker} \\
\hline

\hline

Input &
PDF/DOCX paper, single reference, BibTeX entry, or reference text. &
PDF paper. &
PDF paper. &
ACL-style academic PDF paper (acl.sty, two-column, natbib) or a pre-existing  bib file.  &
ArXiv IDs/URLs, PDFs, LaTeX (.tex), BibTeX (.bib/.bbl), and plain text.\\
\hline
Reference parsing &
Extracts bibliographic fields such as title, authors, journal, year, and DOI. &
Uses a citation recognition model to extract the citation title, authors, year from the reference string. &
Extracts title, authors, DOI, arXiv ID, and the raw citation text. &
Extracts references from PDF,
 or BibTeX and parses authors, titles, years, and venues for later verification. &
Parses titles, authors, years, venues, DOIs, ArXiv IDs, and URLs from the input references.\\
\hline
Main matching method &
Compares title, authors, journal, year, and DOI with records found in scholarly databases. &
Mainly uses title-based fuzzy matching against local databases. &
Uses title, author, DOI, arXiv ID, and database matching to verify references. &
Verifies each reference  based on title , author , year , and cross-database consensus. &
Combines metadata matching with deterministic pre-filters and LLM-assisted deep web search and metadata reverification for suspicious cases.\\
\hline
Databases &
Crossref, Semantic Scholar, and OpenAlex. &
Local ACL Anthology, arXiv, and DBLP databases. &
Crossref, Semantic Scholar, OpenAlex, arXiv, DBLP, PubMed, Europe PMC, ACL Anthology, DOI resolver, and others. &
Semantic Scholar, Crossref, DBLP, and OpenAlex. &
Semantic Scholar, OpenAlex, Crossref, DBLP, ACL Anthology.\\
\hline
Output &
Verified, partial match, mismatch, or suspicious reference. &
Suspicious citation candidates and optionally a highlighted PDF. &
Reference-level status such as verified, not found, or mismatch, with detailed database results. &
JSON and annotated BibTeX output showing flagged references, best matches, and hallucination scores.&
JSON, JSONL, CSV, or text reports with errors, warnings, corrections, and hallucination assessments.\\
\hline
\end{tabular}%
}
\end{table}

\subsection{Evidence from practical testing}
While comparing the documented features of existing tools is important, it is equally essential to evaluate how the tools perform in practical scenarios. We therefore conducted a practical comparison using a manually verified test set of 104 references from three scholarly documents, including 33 problematic references. The test set included incorrect titles, mismatched authors and publication venues, unresolved or incorrect DOIs, unverifiable references, and inconsistencies in publication status. Table 3 summarizes the test documents and the numbers of verified and problematic references for each document. 
\begin{table}[H]
\centering
\small
\renewcommand{\arraystretch}{1.2}
\caption{Summary of the manually verified test documents used for the practical comparison.}
\label{tab:test_documents}

\begin{tabular}{|c|c|c|c|}
\hline
\textbf{Document} & \textbf{Total references} & \textbf{Verified} & \textbf{Hallucinated / problematic} \\
\hline
P1 & 24 & 21 & 3 \\
\hline
P2 & 15 & 11 & 4 \\
\hline
P3 & 65 & 39 & 26 \\
\hline
\textbf{Total} & \textbf{104} & \textbf{71} & \textbf{33} \\
\hline
\end{tabular}
\end{table}

Each tool was tested using the input formats it supported, including full PDFs, individual reference texts, and BibTeX records where possible. The evaluation examined how the tools identified and reported the selected reference problems, as well as the practical limitations observed during testing. Table 4 summarizes the results for each tool.
\begin{table}[htbp]
\centering
\caption{Summary of the practical evaluation results for the reference-checking tools on the manually verified test set.}
\label{tab:tool_performance_summary}

\small
\renewcommand{\arraystretch}{1.3}
\setlength{\tabcolsep}{4pt}

\begin{tabularx}{\linewidth}{
|>{\raggedright\arraybackslash}p{0.15\linewidth}
|>{\centering\arraybackslash}X
|>{\centering\arraybackslash}X
|>{\centering\arraybackslash}X
|>{\centering\arraybackslash}X|
}
\hline

\textbf{Tool} &
\textbf{Correctly detected problematic / hallucinated refs (TP)} &
\textbf{Correctly accepted real / verified refs (TN)} &
\textbf{Real refs incorrectly flagged as problematic (FP)} &
\textbf{Problematic / hallucinated refs missed by the tool (FN)}
\\
\hline

CheckIfExist & 31 & 37 & 34 & 2 \\
\hline
HalluCiteChecker & 18 & 51 & 20 & 15 \\
\hline
Hallucinator & 29  & 43 & 28 & 4 \\
\hline
HalRef & 24 & 18 & 53 & 9  \\
\hline
RefChecker & 32 & 35 & 36 & 1 \\
\hline

\end{tabularx}
\end{table}

The practical comparison showed that the main differences between the tools lie not only in how many problematic references they detect, but also in how often they incorrectly flag valid references. These results illustrate the different strengths and limitations of the tools. RefChecker and CheckIfExist identified more problematic references, but they also produced many false positives. HalluCiteChecker generated fewer false positives but missed more problematic references. Hallucinator showed a more balanced performance, while HalRef provided detailed metadata-related signals but also produced many false positives. General AI text detectors, such as Pangram \parencite{pangram2026}, are not suitable for evaluating hallucinated citations because they do not verify whether references exist or whether their metadata are correct.

Common practical limitations include inaccurate reference extraction from PDF documents, sensitivity to BibTeX formatting, limited database coverage, API restrictions, and the need to configure additional components such as GROBID or LLM-based extractors correctly. Therefore, the output of these tools should not be treated as a final decision. In particular, references labeled as not found, mismatch, warning, or assigned a high hallucination score should be manually checked by comparing the DOI, title, authors, and publication venue.

\section{Conclusion}
Consequently, hallucinated and suspicious references should be regarded as a real and growing challenge for scholarly communication. Recent studies have shown that this problem occurs both across large-scale publication corpora and within conference proceedings. In our study, the evaluated tools successfully identified potentially problematic references and proved useful for initial screening and flagging. However, their performance was not sufficiently reliable to support definitive decisions without further verification. These findings highlight the need for more transparent, multi-source verification systems capable of assessing both the existence and the bibliographic accuracy of references. At the same time, responsible use of generative AI by authors, together with platform-level safeguards, will remain essential for preventing and mitigating this problem.

\section*{Data Availability}

\noindent The manually verified reference-level dataset supporting this study is publicly available in the
\href{https://doi.org/10.5281/zenodo.21457492}{Zenodo repository} \parencite{badalova2026manual}.

\section*{Author Contributions}

\noindent F.B. curated the data, conducted the investigation and formal analysis, implemented the software workflow, validated the results, prepared the visualizations, and drafted and edited the manuscript. P.M. contributed to the conceptualization and methodology, supervised the study, validated the results, and reviewed and edited the manuscript. Both authors approved the final version of the manuscript.

\section*{Competing Interests}

\noindent The authors declare that they have no competing interests.

\section*{Acknowledgement}
The authors of this work were funded by the European Union under the Horizon Europe grant OMINO - Overcoming Multilevel INformation Overload (grant number 101086321, http://ominoproject.eu). Views and opinions expressed are those of the authors alone and do not necessarily reflect those of the European Union or the European Research Executive Agency. Neither the European Union nor the European Research Executive Agency can be held responsible for them. F.B. also acknowledges support from Deutsche Forschungsgemeinschaft (DFG) under grant number MA 3964/15-3, the SocioHub project. 

\printbibliography[title={References}]

\end{document}